\documentclass{ws-procs9x6ujm}
\begin{document}
\title{``B'' is for Bohr}
\author{ULRICH MOHRHOFF}
\address{Sri Aurobindo International Centre of Education\\
Pondicherry 605002, India\\
E-mail: ujm@auromail.net}

\begin{abstract}
It is suggested that the ``B'' in QBism rightfully stands for Bohr. The paper begins by explaining why Bohr seems obscure to most physicists. Having identified the contextuality of physical quantities as Bohr's essential contribution to Kant's theory of science, it outlines the latter, its proper contextuality (human experience), and its decontextualization. In order to preserve the decontextualization achieved by Kant's theory, Bohr seized on quantum phenomena as the principal referents of atomic physics, all the while keeping the universal context of human experience at the center of his philosophy. QBism, through its emphasis on the individual experiencing subject, brings home the intersubjective constitution of objectivity more forcefully than Bohr ever did. If measurements are irreversible and outcomes definite, it is because the experiences of each subject are irreversible and definite. Bohr's insights, on the other hand, are exceedingly useful in clarifying the QBist position, attenuating its excesses, and enhancing its internal consistency.
\end{abstract}

\keywords{Bohr; contextuality; decontextualization; experience; intersubjectivity; objectivity; Kant; QBism.}

\bodymatter
\section{Introduction}\label{ujmsec:intro}
{\leftskip\parindent\emph{I have this ``madly optimistic'' (Mermin called it) feeling that Bohrian--Paulian ideas will lead us to the next stage of physics. That is, that thinking about quantum foundations from their point of view will be the beginning of a new path, not the end of an old one.}\par\hfill--- \emph{Christopher A. Fuchs}\cite{FuchsPaulian}}\par\smallskip\noindent
The beginning of the 21st Century saw the launch of a new interpretation of quantum mechanics, by {Carlton Caves}, {Chris Fuchs}, and {Ruediger Schack}.\cite{CFS2002} Initially conceived as an extended personalist Bayesian theory of probability called {``Quantum Bayesianism},'' it has since been re-branded as ``QBism,'' the term David Mermin\cite{MerminQBnotCop} prefers, considering it ``as big a break with 20th century ways of thinking about science as Cubism was with 19th century ways of thinking about art.'' The big break lies not in the emphasis that the mathematical apparatus of quantum mechanics is a probability calculus---that ought to surprise no one---but in this \emph{plus} a radically subjective Bayesian interpretation of probability \emph{plus} a radically subjective interpretation of the events to which (and on the basis of which) probabilities are assigned.

Recently the referent of the ``B'' in QBism became moot. While Mermin\cite{MerminNow2013} at one time suggested that it should stand for Bruno de Finetti (``Quantum Brunoism''), he now endorses the term Bettabilitarianism\cite{MerminBetterSense} suggested by Chris Fuchs, which was coined by Oliver Wendell Holmes, Jr.:
\begin{quote}
I must not say necessary about the universe\dots. We don't know whether anything is necessary or not. I believe that we can \emph{bet} on the behavior of the universe in its contact with us. So I describe myself as a \emph{bet\,}tabilitarian.\cite{Holmes-Pollock}
\end{quote}
In the present paper, I shall argue that the ``B'' in QBism rightfully stands for Niels Bohr.

The paper is organized as follows. Section~\ref{ujmsec:obscure} explains why Bohr nowadays seems obscure to most physicists.  Section~\ref{ujmsec:contextuality} identifies the contextuality of physical variables as Bohr's essential contribution to Kant's theory of science. (Affinities between Bohr's theory of science and Kant's have been noted by a number of scholars.\cite{Falkenburg2010, Brock2010, dEspagnat2010, Chevalley94, Folse94, Hooker94, Bitbol2010, Cuffaro2010, Honner1982, Kaiser1992, MacKinnon2012}) Section~\ref{ujmsec:decontexting} outlines Kant's theory of science, its own contextuality---human experience---and its decontextualization. Section~\ref{ujmsec:realism-gbu} distinguishes three kinds of realism: the ``good'' (internal) realism of Kant and Bohr, the ``bad'' naive realism recently defended by John Searle,\cite{Searle2004} and the ``ugly'' realism associated with the representative theory of perception. 

For Kant, objectivity meant the possibility of thinking of experiences as experiences of objects. Realizing that in the new field of experience opened up by the quantum theory this possibility no longer exists, Bohr supplemented the object-oriented language of everyday discourse with the language of quantum phenomena. This is discussed in Sec.~\ref{ujmsec:Bohr-o2qp}. Section~\ref{ujmsec:Bohr-cl} explains why Bohr insisted on the need to use (i)~``ordinary language'' or ``plain language'' or ``the common human language'' and (ii)~ ``classical concepts'' or ``the terminology of classical physics'' yet never mentioned ``classical language'' nor ``the language of classical physics'' (i.e., not once in his \emph{Collected Works}). Section~\ref{ujmsec:amplification} addresses the apparent conflict in Bohr's writings between invocations of ``irreversible processes'' and ``irreversible amplification effects'' on the one hand and ``the essential irreversibility inherent in the very concept of observation'' on the other. Section~\ref{ujmsec:years} briefly surveys the philosophically rather barren period between the passing of Niels Bohr and the advent of QBism.

The discussion of QBism begins in Sec.~\ref{ujmsec:qbismWigner}, in which it is argued that by admitting incoherent superpositions of alternatives involving distinct cognitive states, the QBist solutions to ``Wigner's friend''\cite{Wigner61} and similar conundrums overshoot their marks. Section~\ref{ujmsec:qbismShifty} concerns the placement of Bell's shifty split.\cite{Bell90} QBists see only one alternative to placing the Heisenberg cut inside the objective world, to wit, between the objective world and the private experiences in which it originates. There is, however, another alternative, which is to place it between the objective world and the unspeakable domain beyond the reach of our concepts, which only becomes speakable by saying in ordinary language ``what we have done and what we have learned'' [BCW\,7:349].%
\footnote{In what follows, BCW (followed by volume and page number) refers to the \emph{Collected Works} of Niels Bohr.\cite{BCW}}
This, I content, is where it should be placed, and where to all intents and purposes it was placed by Bohr.

The concluding section deals with how QBism relates to Bohr, beginning with certain misreadings of Bohr that QBists share with the majority of current interpreters of quantum mechanics. My final verdict is that QBism, through its emphasis on the individual experiencing subject, brings home the intersubjective constitution of our common external world more forcefully than Bohr ever did. (The time wasn't ripe for this then. Perhaps it is now.) Bohr's insights, on the other hand, are exceedingly useful in clarifying the QBist position, attenuating its excesses, and enhancing its internal consistency.

\section{Why Bohr seems obscure}\label{ujmsec:obscure}
{\leftskip\parindent\emph{As a philosopher Niels Bohr was either one of the great visionary figures of all time, or merely the only person courageous enough to confront head on, whether or not successfully, the most imponderable mystery we have yet unearthed. --- N. David Mermin}\cite{MerminBoojums}\par}\smallskip\noindent
Today, Bohr seems obscure to most physicists. Catherine Chevalley\cite{Chevalley99} has identified three reasons for this deplorable situation. The first is that Bohr's mature views, which ``remained more or less stable at least over the latter thirty years of Bohr's life''\cite{Hooker1972} (i.e., since at least 1932), have come to be equated with one variant or another of the Copenhagen interpretation. The latter only emerged in the mid-1950's, in response to David Bohm's hidden-variables theory and the Marxist critique of Bohr's alleged idealism, which had inspired Bohm. The second reason is that Bohr's readers will usually not find in his writings what they expected to find, while they will find a number of things that they did not expect. What they expect is a take on problems arising in the context of $\Psi$-ontology---the spurious reification of a probability calculus---such as the problem of objectification or the quantum-to-classical transition. What they find instead is discussions of philosophical issues such as the meanings of ``objectivity,''  ``truth,'' and ``reality'' and the roles of language and communication. The third reason is that the task of making sense of quantum mechanics is seen today as one of grafting a metaphysical narrative onto a mathematical formalism, in a language that is sufficiently vague philosophically to be understood by all and sundry. For Bohr, as also for Werner Heisenberg and Wolfgang Pauli, the real issues lay deeper. They judged that the conceptual difficulties posed by quantum mechanics called in question the general framework of contemporary thought, its concepts, and its criteria of consistency. 

\section{Contextuality}\label{ujmsec:contextuality}
It was Immanuel Kant, the most important philosopher of the modern era, who first demonstrated that it was possible to provide a scientific theory with much stronger justification than mere empirical adequacy. The kind of argument inaugurated by him to this end begins by assuming that a certain proposition \textbf{p} is true, and then shows that another proposition \textbf{q}, stating a precondition for the truth of \textbf{p}, must also be true: if \textbf{q} were not true, \textbf{p} could not be true. For his immediate purpose the relevant proposition \textbf{p} was that empirical knowledge is possible, and the corresponding proposition \textbf{q} was that certain universal laws must hold.

``Reason,'' Kant wrote, ``must approach nature with its principles in one hand \dots\ and, in the other hand, the experiments thought out in accordance with these principles.'' The concepts in terms of which reason's principles are formulated owe their meanings to our cognitive faculties of intuition%
\footnote{The German original, \emph{Anschauung}, covers both visual perception and visual imagination.}
and thought. They allow us to ask meaningful questions, and to make sense of the answers we obtain by experiment and observation: ``what reason would not be able to know of itself and has to learn from nature, it has to seek in the latter'' but it has to do this ``in accordance with what reason itself puts into nature''.\cite{KantCPR2} What Kant did not anticipate was that experiments would come to play the same constitutive role as our cognitive faculties do in defining the terms of our discourse with nature. The insight that certain questions are \emph{contextual}---that they have no answers unless their answers are elicited by actual experiments---is due to Bohr. To hitch the definition of physical quantities to the experimental conditions under which they are observed, is Bohr's ground-breaking contribution to Kant's theory of science.\cite{Bitbol98} It has, moreover, been spectacularly borne out by the no-go theorems of John Bell,\cite{Bell64} Simon Kochen and Ernst Specker,\cite{Kochen-Specker} and Alexander Klyachko and coworkers.\cite{Klyachko08}

\section{Decontextualization}\label{ujmsec:decontexting}
Bohr's contextuality, however, was not the first to play a role in natural philosophy. From the end of the 17th century onwards, it was widely accepted by philosophers that objects existed relative to a context, to wit, human experience. By placing the subject of empirical science squarely into the context of human experience, Kant dispelled many qualms that had been shared by thinkers at the end of the 18th century---qualms about the objective nature of geometry, about the purely mathematical nature of Newton's theory, about the unintelligibility of action at a distance, about Galileo's principle of relativity, to name a few. 

Concerning the laws of geometry, which apply to objects constructed by us in the space of our imagination, the question was why they should also apply to the physical world. Kant's answer was that they apply to objects perceived as well as to objects imagined because visual perception and visual imagination share the same space.%
\footnote{\label{note:Kant1}It is noteworthy that Kant's argument applies, not to Euclidean geometry specifically, which was the only geometry known in Kant's time, but to geometry in general, and thus to whichever geometry is best suited to formulating the laws of physics. It has even been said that Kant's theory of science set in motion a series of re-conceptualizations of the relationship between geometry and physics that eventuated in Einstein's theories of relativity.\cite{Friedman2009}}
As to the mathematical nature of Newtonian mechanics, it was justified, not by the Neo-Platonic belief that the book of nature was written in mathematical language, but by its being a precondition of scientific knowledge. What made it possible to conceive of appearances as aspects of an objective world was the mathematical regularities that obtain between them. Newton's refusal to explain action at a distance was similarly justified, inasmuch as the only intelligible causality available to us consists in lawful mathematical relations between phenomena: for the Moon to be causally related to the Earth was for the Moon to stand in a regular mathematical relation to the Earth. As to the principle of relativity, ditto: lawful mathematical relations only exist between phenomena, and thus only between objects or objective events, but never between a particular phenomenon and space or time itself.%
\footnote{\label{note:Kant2}Here, too, it would be an anachronism to argue that Kant singled out Galilean relativity, which was the only relativity known in his time. His argument holds for every possible principle of relativity, including Einstein's.}

Kant's premise was that ``space and time are only forms of sensible intuition, and therefore only conditions of the existence of the things {as appearances}.'' It follows
\begin{quote}
that we have no concepts of the understanding and hence no elements for the cognition of things except insofar as an intuition can be given corresponding to these concepts, consequently that we can have cognition of no object as a thing in itself, but only insofar as it is an object of sensible intuition, i.e. as an appearance; from which follows the limitation of all even possible speculative cognition of reason to mere objects of \emph{experience}. Yet \dots\ even if we cannot \emph{cognize} these same objects as [i.e., \emph{know} them to be] things in themselves, we at least must be able to \emph{think} them as things in themselves. For otherwise there would follow the absurd proposition that there is an appearance without anything that appears.\cite{KantCPR3}
\end{quote}
Before Kant, there appears to have been no philosopher who did not have a correspondence theory of truth, and who did not think of the relation of sense impressions to the external world as a relation of similarity. Kant was the first to show that the predictive success of a scientific theory does not have to be attributed to some empirically inaccessible correspondence between the structure of the theory and the structure of the real world. Needless to say, this had to be done without calling into question the objectivity of the theory, i.e., in a way that allowed people to think of phenomena as appearances of things ``out there.'' We must be able to \emph{decontextualize} the objective world, to forget that it depends on us. And if there is only the single universal context of human experience, this is easily done. We are free to think of perceived objects as faithful representations of real objects (things in themselves), free to forget that the apparently mind-independent system of objects ``out there'' was a mental construct, and that the concepts that were used in its construction are meaningless outside the context of human experience.

\section{Realism good, bad, and ugly}\label{ujmsec:realism-gbu}
In an essay written during the last year of his life,\cite{SchrWhatIsReal} Erwin Schr\"odinger expressed his astonishment at the fact that despite ``the absolute hermetic separation of my sphere of consciousness'' from everyone else's, there was ``a far-reaching structural similarity between certain parts of our experiences, the parts which we call external; it can be expressed in the brief statement that we all live in the same world.'' This similarity, Schr\"odinger avowed, was ``not rationally comprehensible. In order to grasp it we are reduced to two irrational, mystical hypotheses,'' one of which%
\footnote{The alternative hypothesis, which he endorsed, was ``that we are all really only various aspects of the One''\cite{SchrWhatIsReal}: the multiplicity of minds ``is only apparent, in truth there is only one mind. This is the doctrine of the Upanishads. And not only of the Upanishads''.\cite{SchrLifeMindMatterAPOM} The One in question is the ultimate subject, from which we are separated by a veil of self-oblivion. The same veil (according to the Upanishads) also prevents us from perceiving the ultimate object, as well as its identity with the ultimate subject. If ``to Western thought this doctrine has little appeal,'' it is because our science ``is based on objectivation, whereby it has cut itself off from an adequate understanding of the Subject of Cognizance, of the mind.'' For Schr\"odinger,\cite{SchrLifeMindMatterPO} this was ``precisely the point where our present way of thinking does need to be amended, perhaps by a bit of blood-transfusion from Eastern thought.''}
is ``the so-called {hypothesis of the real external world}.'' Schr\"odinger left no room for uncertainty about what he thought of this hypothesis. To invoke ``the existence of a real world of bodies which are the causes of sense impressions and produce roughly the same impression on everybody \dots\ is not to give an explanation at all; it is simply to state the matter in different words. In fact, it means laying a completely useless burden on the understanding.'' It means uselessly translating the statement ``everybody agrees about something'' into the  statement ``there exists a real world which causes everybody's agreement.'' Instead of explaining the fact expressed by the first statement, the second merely reinforces its incomprehensibility, for the relation between this real world and those aspects of our experiences about which there is agreement, is something we cannot know. The causal relations we know are internal to those of our experiences about which we agree.

In ancient and medieval philosophy, to \emph{be} was either to be a substance or to be a property of a substance. Substance was self-existent; everything else depended for its existence on a substance. With Descartes, the human conscious subject assumed the role of substance: to \emph{be} became either to be a subject or to exist as a representation for a subject. Thus was born the representative theory of perception. In the eyes of philosopher John Searle,\cite{Searle2004a} the move from the older view that ``we really perceive real objects'' to the view that we only perceive sense impressions was ``the greatest single disaster in the history of philosophy over the past four centuries.'' A disaster it was indeed, not least because it continues to muddy the scientific waters when it comes to sensory perception.%
\footnote{\label{note:sacop}The standard scientific account of perception begins by positing the mind-independent existence of a real world. Objects in this world are said to emit light or sound waves, which are said to stimulate peripheral nerve endings (retinas or ear drums). The stimulated nerves are said to send signals to the brain, where neural processes are said to give rise to perceptual experience. The trouble with this account is not simply that no one has any idea how to bridge the ``explanatory gap''\cite{Levine2001} between objective brain processes and subjective experiences. The trouble is that no one has any idea how we could have information about what happens or exists in that real world. While the standard scientific account of perception begins by invoking events in that world, it leads to the conclusion that we have access only to our experiences, and that there is no way we could know anything about what happens or exists in that world.}

The representative theory of perception poses this dilemma: either the gap between representations and the objects they are supposed to represent can never be bridged, or the world is reduced to representations. Either science deals with objects in the real world, in which case we have no justifiable idea of how we come to have representations, or it deals with representations, in which case we have no justifiable knowledge of the real world. Transcendental philosophy, inaugurated by Kant and continued in the 20th century by Edmund Husserl,\cite{HusserlEssential} emerged as a critique of the representative theory. In an attempt to defend the older, direct realism, Searle has invoked the fact that we are able communicate with other human beings, using publicly available meanings in a public language. For this to work, he argued,\cite{Searle2004b} we have to assume common, publicly available objects of reference: 
\begin{quote}
So, for example, when I use the expression ``this table'' I have to assume that you understand the expression in the same way that I intend it. I have to assume we are both referring to the same table, and when you understand me in my utterance of ``this table'' you take it as referring to the same object you refer to in this context in your utterance of ``this table.''
\end{quote}
The implication then is that 
\begin{quote}
you and I share a perceptual access to one and the same object. And that is just another way of saying that I have to presuppose that you and I are both seeing or otherwise perceiving the same public object.  But that public availability of that public world is precisely the direct realism that I am here attempting to defend.
\end{quote}
Searle points out that his argument is transcendental in Kant's sense. Here \textbf{p} is the assumption that we are able to communicate with each other in a public language, and \textbf{q} is the conclusion that there must be publicly available objects in a public world about which we can communicate in a public language. The actual implication of his argument, however, is the agreement between our respective ``spheres of consciousness''---between what exists for me, in my experience, and what exists for you, in your experience---which so astonished Schr\"odinger. It allows us to communicate with each other \emph{as if} direct realism were true. What Searle succeeds in defending against the ``ugly'' representative realism is not the ``bad'' direct realism but the ``good'' \emph{internal realism}  inaugurated by Immanuel Kant and defended (among others) by Hilary Putnam and Bernard d'Espagnat.%
\footnote{Putnam assumed the existence of a mind-independent real world but  insisted that it does not dictate its own descriptions to us: ``talk of ordinary empirical objects is not talk of things-in-themselves but only talk of things-for-us''\cite{Putnam81}; ``we don't know what we are talking about when we talk about `things in themselves'\,''\cite{Putnam87}. D'Espagnat,\cite{d'Espagnat_VR} for his part, stressed the necessity of distinguishing between an empirically inaccessible veiled reality and an intersubjectively constructed objective reality.}

The key role that language plays in establishing the rationally incomprehensible correspondence between the ``external parts'' of our internal experiences, has also been emphasized by Schr\"odinger:
\begin{quote}
What does establish it is \emph{language}, including everything in the way of expression, gesture, taking hold of another person, pointing with one's finger and so forth, though none of this breaks through that inexorable, absolute division between spheres of consciousness.\cite{SchrWhatIsReal}
\end{quote}

\section{Bohr: from objects to quantum phenomena}\label{ujmsec:Bohr-o2qp}
The hallmark of empirical knowledge is objectivity. To Kant, objectivity meant the possibility of thinking of appearances as experiences of \emph{objects}. His inquiry into the preconditions of empirical science was therefore an inquiry into the conditions that make it possible to organize sense impressions into (identifiable) objects. Yet in the new field of experience opened up by the quantum theory, this possibility no longer seemed to exist. As Schr\"odinger\cite{SchrNGSH} wrote,
\begin{quote}
Atoms---our modern atoms, the ultimate particles---must no longer be regarded as identifiable individuals. This is a stronger deviation from the original idea of an atom than anybody had ever contemplated. We must be prepared for anything.
\end{quote}
For the present-day physicist, it is not easy to understand the bewilderment that the founders and their contemporaries experienced in the early days of the quantum theory:
\begin{quote}
All the verities of the preceding two centuries, held by physicists and ordinary people alike, simply fell apart---collapsed. We had to start all over again, and we came up with something that worked just beautifully but was so strange that nobody had any idea what it meant except Bohr, and practically nobody could understand him. So naturally we kept probing further, getting to smaller and smaller length scales, waiting for the next revolution to shed some light on the meaning of the old one.\cite{Mermin_epochs}
\end{quote}
That revolution never came. Quantum mechanics works as beautifully in the nucleus as it does in the atom; and it works as beautifully in the nucleon as it does in the nucleus, seven or eight orders of magnitude below the level for which it was designed. (It also works beautifully many orders of magnitude above that level, as for example in a superconductor.) It is therefore past time to try more seriously to understand what Bohr had been trying to drive home.

``Without sensibility no object would be given to us,'' Kant wrote,\cite{KantCPR5} ``and without understanding none would be thought.'' And again: ``we have no concepts of the understanding \dots\ except insofar as an intuition can be given corresponding to these concepts''.\cite{KantCPR3} Bohr could not have agreed more, insisting as he did that meaningful physical concepts have not only mathematical but also visualizable content. Such concepts are associated with pictures, like the picture of a particle following a trajectory or the picture of a wave propagating in space. In the classical theory, a single picture could accommodate all of the properties a system can have. When quantum theory came along, that all-encompassing picture fell apart. Unless certain experimental conditions obtained, it was impossible to picture the electron as following a trajectory (which was nevertheless a routine presupposition in setting up Stern--Gerlach experiments and in interpreting cloud-chamber photographs), and there was no way in which to apply the concept of position. And unless certain other, incompatible, experimental conditions obtained, it was impossible to picture the electron as a traveling wave (which was nevertheless a routine presupposition in interpreting the scattering of electrons by crystals), and there was no way in which to apply the concept of momentum.

Bohr settled on the nexus between pictures, physical concepts, and experimental arrangements as key to ``the task of bringing order into an entirely new field of experience''.\cite{BohrSchilpp} If the visualizable content of physical concepts cannot be described in terms of compatible pictures, it has to be described in terms of something that \emph{can} be so described, and what can be so described are the experimental conditions under which the incompatible physical concepts can be employed. What distinguishes these experimental conditions from the quantum systems under investigation is their accessibility to direct sensory experience.

What Bohr added to Kant's theory of science was his insight that empirical knowledge was not necessarily limited to what is \emph{directly} accessible to our senses, and that, therefore, it does not have to be \emph{solely} a knowledge of sense impressions organized into objects. It can also be a knowledge of objects that are \emph{not} objects of sensible intuition but instead are constituted by experimental conditions, which \emph{are} sense impressions organized into objects. This is why ``the objective character of the description in atomic physics depends on the detailed specification of the experimental conditions under which evidence is gained'' [BCW\,10:215]. Quantum mechanics does not do away with objects of sensible intuition but supplements them with \emph{quantum phenomena}:
\begin{quote}
all unambiguous interpretation of the quantum mechanical formalism involves the fixation of the external conditions, defining the initial state of the atomic system concerned and the character of the possible predictions as regards subsequent observable properties of that system. Any measurement in quantum theory can in fact only refer either to a fixation of the initial state or to the test of such predictions, and it is first the combination of measurements of both kinds which constitutes a well-defined phenomenon. [BCW\,7:312]
\end{quote}

\section{Bohr: concepts and language}\label{ujmsec:Bohr-cl}
The transition in Bohr's thinking from the familiar epistemology of objects to an epistemology supplemented with quantum phenomena resulted from his insight that ``the facts which are revealed to us by the quantum theory \dots\ lie outside the domain of our ordinary forms of perception'' [BCW\,6:217]. As early as 1922, Bohr opined that the difficulties physicists were facing at the time were ``of such a kind that they hardly allow us to hope, within the world of atoms, to implement a description in space and time of the kind corresponding to our usual sensory images'' [BCW\,10:513--514]. By 1926, the mature (non-relativistic) theory was in place, and by 1929 Bohr's thoughts had gelled into what to my mind remains the most astute understanding of quantum mechanics to date. 

In his writings of that year, abundant reference is made to ``our (ordinary) forms of perception,'' time and space. As in: quantum theory has ``justified the old doubt as to the range of our ordinary forms of perception when applied to atomic phenomena'' [BCW\,6:209]; ``at the same time as every doubt regarding the reality of atoms has been removed, \dots we have been reminded in an instructive manner of the natural limitation of our forms of perception'' [BCW\,6:237]. This limitation was ``brought to light by a closer analysis of the applicability of the basic physical concepts in describing atomic phenomena'' [BCW\,6:242]. That is to say, the natural limitation of our forms of perception both implies and is implied by a natural limitation of the applicability of our basic physical concepts, which is a consequence of the uncertainty relations. Yet ``in spite of their limitation, we can by no means dispense with those forms of perception which colour our whole language and in terms of which all experience must ultimately be expressed'' [BCW\,6:283]. In other words, the conceptual framework of quantum physics is the same as that of classical physics, the difference being that in quantum physics its applicability is limited.

``When speaking of a conceptual framework,'' Bohr wrote, ``we merely refer to an unambiguous logical representation of relations between experiences'' [BCW\,10:84]. In Bohr's time and the cultural environment in which he lived, Kant's theory of science still exercised considerable influence. There can be little doubt that the unambiguous logical representation of relations between experiences that Bohr had in mind, was in all important respects the conceptual framework staked out by Kant, providing the general structure of an object-oriented language. 

In Kant's theory of science, the relevant relations between experiences fall under the logical categories of substance, causality, and interaction. The logical relation between a (logical) subject and a predicate makes it possible for us to think of a particular nexus of perceptions as the properties of a \emph{substance}, connected to it as predicates are connected to a subject. It makes it possible for me to think of perceptions as connected not by me, in my experience, but in an object ``out there'' in the public world. The logical relation between antecedent and consequent (if \dots\ then\dots) makes it possible for us to think of what we perceive at different times as properties of substances connected by \emph{causality}. It makes it possible for me to think of asynchronous perceptions as connected not merely in my experience but also objectively, by a causal nexus ``out there.'' And the category of community or reciprocity, which Kant associated with the disjunctive relation (either\dots\ or\dots), makes it possible for us to think of what we perceive in different locations as properties of substances connected by a \emph{reciprocal action}. It makes it possible for me to think of simultaneous perceptions as connected not only in my experience but objectively. (Kant thought that by establishing a reciprocal relation, we establish not only an objective spatial relation but also an objective relation of simultaneity.)

But if we are to be able to think of perceptions as properties of substances, or as causally connected, or as affecting each other, the connections must be regular. For perceptions to be perceptions of a particular kind of thing (say, an elephant), they must be connected in an orderly way, according to a concept denoting a lawful concurrence of perceptions. For perceptions to be causally connected, like (say) lightning and thunder, they must fall under a causal law, according to which one perception necessitates the subsequent occurrence of another. (By establishing a causal relation falling under a causal law, we also establish an objective temporal relation.) And for perceptions to be reciprocally connected, like (say) the Earth and the Moon, they must affect each other according to a reciprocal law, such as Newton's law of gravity. It is through lawful connections in the ``manifold of appearances'' that we are able to think of appearances as perceptions of a self-existent system of causally evolving (and thus re-identifiable) objects, from which we, the experiencing subjects, can remove ourselves. 

Even in a field of experience in which the concepts required to bundle sense impressions into objects cannot be applied, one has to rely on the common object-oriented language. Where one cannot speak of objects, one has to speak of quantum phenomena, i.e., of experimental arrangements and results indicated by measuring instruments: 
\begin{quote}
The argument is simply that by the word ``experiment'' we refer to a situation where we can tell others what we have done and what we have learned and that, therefore, the account of the experimental arrangement and of the results of the observations must be expressed in unambiguous language with suitable application of the terminology of classical physics. [BCW\,7:349]
\end{quote}
Two expressions are significant here: ``unambiguous language'' and ``the terminology of classical physics.'' Presently (August 2019) a combined Google search for ``Bohr'' and ``classical language'' (the latter term including the quotes) yields nearly 5,000 results, while a Google search for ``Bohr'' and ``language of classical physics'' yields more than 24,000 results. By contrast, searching the 13 volumes of the \emph{Complete Works} of Niels Bohr does not yield a \emph{single} occurrence of either ``classical language'' or ``language of classical physics.'' It is the ubiquity in the secondary literature of these latter expressions that is chiefly responsible for the widespread misconceptions about Bohr's thinking.

While Bohr insisted on the use of classical \emph{concepts} (or, the terminology of classical physics, or simply ``elementary physical concepts'' [BCW\,7:394]), the \emph{language} on the use of which he insisted was ``ordinary language'' [BCW\,7:355], ``plain language'' [BCW\,10:159], the ``common human language'' [BCW\,10:157--158], or the ``language common to all'' [10:xxxvii]. A distinction must therefore be drawn between the role that classical concepts played in Bohr's thinking and the role that was played by the common human language. The common human language is the object-oriented language of everyday discourse, while classical concepts are not proprietary to classical physics but denote attributes that owe their meanings to our forms of perception, such as position, orientation, and the ones that are defined in terms of invariances under spacetime transformations.

One day during tea at his institute, Bohr was sitting next to Edward Teller and Carl Friedrich von Weizs\"acker. Von Weizs\"acker\cite{vW_Structure} recalls that when Teller suggested that ``after a longer period of getting accustomed to quantum theory we might be able after all to replace the classical concepts by quantum theoretical ones,'' Bohr listened, apparently absent-mindedly,  and said at last: ``Oh, I understand. We also might as well say that we are not sitting here and drinking tea but that all this is merely a dream.'' If we are dreaming, we are unable to tell others what we have done and what we have learned. Therefore
\begin{quote}
it would be a misconception to believe that the difficulties of the atomic theory may be evaded by eventually replacing the concepts of classical physics by new conceptual forms. \dots the recognition of the limitation of our forms of perception by no means implies that we can dispense with our customary ideas or their direct verbal expressions when reducing our sense impressions to order. [BCW\,6:294]
\end{quote}
Or, as Heisenberg put it,\cite{Heisenberg_PP56} ``[t]here is no use in discussing what could be done if we were other beings than we are.''%
\footnote{Heisenberg thought it possible that the forms of perception of other beings, and hence their concepts, could be different from ours: our concepts ``may belong to the species `man,' but not to the world as independent of men''.\cite{Heisenberg_PP91}} 
Bohr's claim that the ``classical language'' (i.e., plain language supplemented with the terminology of classical physics) was indispensable, has also been vindicated by subsequent developments in particle physics:
\begin{quote}
This [claim] has remained valid up to the present day. At the \emph{individual} level of clicks in particle detectors and particle tracks on photographs, all measurement results have to be expressed in classical terms. Indeed, the use of the familiar physical quantities of length, time, mass, and momentum-energy at a subatomic scale is due to an extrapolation of the language of classical physics to the non-classical domain.\cite{Falkenburg2007-162}
\end{quote}
It is therefore an irony that Bohr, seeing Kant as arguing for the necessary validity and unlimited reach of classical concepts, regarded complementarity as an alternative to Kant's theory of science, thus drawing the battle lines in a way which put Kant and himself on opposing sides. Just as Kant did not argue for the universal validity of Euclidean geometry \emph{in particular} (see Note~\ref{note:Kant1}), nor for Galilean relativity \emph{in particular} (see Note~\ref{note:Kant2}), so his arguments did not, in effect, establish the \emph{unlimited reach} of classical concepts. As his arguments merely established the validity of whichever geometry, and whichever principle of relativity, was convenient, so they established the necessary validity of classical concepts but not their unlimited reach. What Kant did not anticipate was the possibility of empirical knowledge that did \emph{not} involve the organization of sense impressions into objects---an empirical knowledge that, while being obtained \emph{by means of} sense impressions organized into objects, was not a knowledge \emph{of} sense impressions organized into objects. 

\section{Bohr and the irreversibility of measurements}\label{ujmsec:amplification}
If the terminology of quantum phenomena is used consistently, then nothing---at any rate, nothing we know how to think about---happens between ``the fixation of the external conditions, defining the initial state of the atomic system concerned'' and ``the subsequent observable properties of that system'' [BCW\,7:312]. 
Any story purporting to detail a course of events in the interval between a system preparation and a subsequent observation is inconsistent with ``the essential wholeness of a quantum phenomenon,'' which ``finds its logical expression in the circumstance that any attempt at its subdivision would demand a change in the experimental arrangement incompatible with its appearance'' [BCW\,10:278]. What, then, are we to make of the following passages [emphases added]?
\begin{quote}
[E]very well-defined atomic phenomenon is closed in itself, since its observation implies a permanent mark on a photographic plate \emph{left by the impact of an electron} or similar recordings \emph{obtained by suitable amplification devices of essentially irreversible functioning}. [BCW\,10:89]

\smallskip Information concerning atomic objects consists solely in the marks they make on these measuring instruments, as, for instance, a spot \emph{produced by the impact of an electron on a photographic plate} placed in the experimental arrangement. The circumstance that such marks are \emph{due to irreversible amplification effects} endows the phenomena with a peculiarly closed character pointing directly to the irreversibility in principle of the very notion of observation. [BCW\,10:120]

\smallskip In this connection, it is also essential to remember that all unambiguous information concerning atomic objects is derived from the permanent marks---such as a spot on a photographic plate, \emph{caused by the impact of an electron}---left on the bodies which define the experimental conditions. Far from involving any special intricacy, the\emph{ irreversible amplification effects on which the recording of the presence of atomic objects rests} rather remind us of the essential irreversibility inherent in the very concept of observation. [BCW\,7:390; BCW\,10:128]
\end{quote}
If a well-defined atomic phenomenon is closed, how can something happen between the fixation of the external conditions and a permanent mark on a photographic plate? Does not the interposition of the impact of an electron and/or of subsequent amplification effects amount to a subdivision of the phenomenon in question? 

Ole Ulfbeck and Aage Bohr\cite{UlfbeckBohr} have shed light on this issue. For them,  clicks in counters are ``events in spacetime, belonging to the world of experience.'' While clicks can be classified as electron clicks, neutron clicks, etc., ``there are no electrons and neutrons on the spacetime scene'' and ``there is no wave function for an electron or a neutron but [only] a wave function for electron clicks and neutron clicks.''  ``[T]here is no longer a particle passing through the apparatus and producing the click. Instead, the connection between source and counter is inherently non-local.'' The key to resolving the issue at hand is that each click has an ``onset''---``a beginning from which the click evolves as a signal in the counter.'' This onset 
\begin{quote}
 has no precursor in spacetime and, hence, does not belong to a chain of causal events. In other words, the onset of the click is not the effect of something, and it has no meaning to ask how the onset occurred\dots. [T]he occurrence of genuinely fortuitous clicks, coming by themselves, is recognized as the basic material that quantum mechanics deals with\dots. [T]he wave function enters the theory not as an independent element, but in the role of encoding the probability distributions for the clicks\dots. [T]he steps in the development of the click, envisaged in the usual picture, are not events that have taken place on the spacetime scene\dots. [T]he downward path from macroscopic events in spacetime, which in standard quantum mechanics continues into the regime of the particles, does not extend beyond the onsets of the clicks.
\end{quote}
If irreversible amplification effects---the steps in the development of the click---only occur ``in the usual picture,'' then they neither modify nor subdivide the quantum phenomenon in which---through an illegitimate extension of the object-oriented language of classical physics---they are said to occur.

For Niels Bohr, ``the physical content of quantum mechanics is exhausted by its power to formulate statistical laws governing observations obtained under conditions specified in plain language'' [BCW\,10:159]. A quantum phenomenon thus has a statistical component, which correlates events in the world of experience. The so-called irreversible amplification effects belong to this statistical component. The unmediated step from the source to the onset of the click, and the subsequent unmediated steps in the development of the click, are steps in a gazillion of alternative sequences of possible outcomes of \emph{unperformed} measurements, and unperformed measurements do not affect the essential wholeness of a quantum phenomenon.

Niels Bohr, moreover, strongly cautioned against the terminology of ``disturbing a phenomenon by observation'' and of ``creating physical attributes to objects by measuring processes'' [BCW\,7:316]. If there is nothing to be disturbed by observation, if even the dichotomy of objects and attributes created for them by measuring processes is unwarranted, then it is not just the measured property that is constituted by the experimental conditions under which it is observed; it is the quantum system itself that is so constituted. Recently this point was forcefully made by Brigitte Falkenburg in her monograph \emph{Particle Metaphysics}\cite{Falkenburg2007-205f}:
\begin{quote}
[O]nly the experimental context (and our ways of conceiving of it in classical terms) makes it possible to talk in a sloppy way of \emph{quantum objects}\dots. Bare quantum ``objects'' are just bundles of properties which underlie superselection rules and which exhibit non-local, acausal correlations\dots. They seem to be Lockean empirical substances, that is, collections of empirical properties which constantly go together. However, they are only individuated by the experimental apparatus in which they are measured or the concrete quantum phenomenon to which they belong\dots. They can only be individuated as context-dependent quantum \emph{phenomena}. Without a given experimental context, the reference of quantum concepts goes astray. In this point, Bohr is absolutely right up to the present day.
\end{quote}

In the following passages [emphases added], Bohr goes beyond invoking irreversible amplification effects, apparently arguing that the quantum features involved in the atomic constitution of a measurement apparatus (or the statistical element in its description) can be neglected because the relevant parts of a measurement apparatus are sufficiently large and heavy.
\begin{quote}
In actual experimentation this demand [that the experimental arrangement as well as the recording of observations be expressed in the common language] is met by the specification of the experimental conditions by means of bodies like diaphragms and photographic plates \emph{so large and heavy that the statistical element in their description can be neglected}. The observations consist in the recording of permanent marks on these instruments, and the fact that the amplification devices used in the production of such marks involves essentially irreversible processes presents no new observational problem, but merely stresses the element of irreversibility inherent in the definition of the very concept of observation. [BCW\,10:212]

\smallskip In actual physical experimentation this requirement [that we must employ common language to communicate what we have done and what we have learned by putting questions to nature in the form of experiments] is fulfilled by using as measuring instruments rigid bodies like diaphragms, lenses, and photographic plates \emph{sufficiently large and heavy to allow an account of their shape and relative positions and displacements without regard to any quantum features inherently involved in their atomic constitution}\dots. The circumstance that [recordings of observations like the spot produced on a photographic plate by the impact of an electron] involve essentially irreversible processes presents no special difficulty for the interpretation of the experiments, but rather stresses the irreversibility which is implied in principle in the very concept of observation. [BCW\,10:165]
\end{quote}
How can the size and weight of a measuring device justify
\begin{itemize}
\item[---] the irreversibility in principle of the very notion of observation [BCW\,10:120],
\item[---] the essential irreversibility inherent in the very concept of observation [BCW\,7:390; BCW\,10:128],
\item[---] the irreversibility which is implied in principle in the very concept of observation [BCW\,10:165], or
\item[---] the element of irreversibility inherent in the definition of the very concept of observation [BCW\,10:212]?
\end{itemize}
The only irreversibility that can justify the irreversibility of observations is the incontestable irreversibility of human sensory experience. 
For Bohr, ``the emphasis on the subjective character of the idea of observation [was] essential'' [BCW\,10:496]. If, as he insisted, the description of atomic phenomena nevertheless has ``a perfectly objective character,'' it was ``in the sense that no explicit reference is made to any \emph{individual} observer and that therefore \dots\ no ambiguity is involved in the communication of information'' [BCW\,10:128, emphasis added]. It was never in the sense that no reference was made to the community of communicating observers or to the incontestable irreversibility of their experiences.

Like Kant, Bohr was concerned with the possibility of an objective knowledge of phenomena (or appearances, or experiences). Kant primarily addressed the possibility of organizing appearances into a world of objects. This involved concepts that allowed the individual not only to think of appearances as a world of external objects but also to communicate with others about a world of external objects. Bohr primarily addressed the requisite possibility of communication. This involved the use of concepts that allowed the individual not only to communicate with others about a world of external objects but also to think of appearances as a world of external objects. Bohr and Kant were on the same page.

What distinguishes the objectivity that could be achieved in the age of Kant from the objectivity that can be achieved in the age of quantum mechanics is that a complete elision of the subject is no longer feasible. Having asserted that ``we can have cognition of no object as a thing in itself, but only insofar as it is an object of sensible intuition, i.e. as an appearance,'' Kant (as we have seen) went on to affirm that
\begin{quote}
even if we cannot \emph{cognize} these same objects as things in themselves, we at least must be able to \emph{think} them as things in themselves. For otherwise there would follow the absurd proposition that there is an appearance without anything that appears.\cite{KantCPR3}
\end{quote}
If the only relevant context is human experience, or if the reach of human sensory experience is unlimited (as classical physics takes it to be), the elision of the subject can be achieved. It is possible to transmogrify calculational tools into objective physical processes with some measure of consistency.\cite{Mermin_badhabit} But if the experimental context is relevant as well, or if the reach of human sensory experience is limited, the elision of the subject is a lost cause, and so is the transmogrification of calculational tools into physical mechanism or natural processes. If one nevertheless wants to establish the irreversibility of measurements and the definiteness of outcomes without invoking the definiteness and irreversibility of human sensory experience, one has no choice but to invoke such ``macroscopic'' features as the size and weight of a measurement apparatus. This (to appropriate a well-known passage by John Bell),
\begin{quote}
is like a snake trying to swallow itself by the tail. It can be done---up to a point. But it becomes embarrassing for the spectators even before it becomes uncomfortable for the snake.\cite{Bell90}
\end{quote}
In the words of Bernard d'Espagnat\cite{dEspagnat2010}, quantum mechanics practically compels us
\begin{quote}
to adopt the idea that was, in fact, at the very core of Kantism and constitutes its truly original contribution to philosophical thinking, to wit, the view that things and events, far from being elements of a ``reality per se,'' are just phenomena, that is, elements of our experience.
\end{quote}
So what were Bohr's intentions in invoking, in lieu of the irreversibility of human experience, the size and weight of the measurement apparatus? Was it in order to appease the na\"{\i}ve realistic inclinations of lesser minds? Or was it in order to indicate the price one had to pay (but which he personally was not willing to pay) for exorcising every possible reference to human sensory experience? My suggested answer is ``yes'' to both questions.

\section{The intervening years}\label{ujmsec:years}
In 1932 John von Neumann\cite{vN1932} developed the mathematical part of the quantum theory into an autonomous formal language. In doing so he turned the theory into a mathematical formalism that was in need of a physical interpretation. In the 1950s, interpreting quantum mechanics began to turn into a growth industry. First David Bohm presented his hidden-variables interpretation,\cite{Bohm1952} then Hugh Everett put forward his relative-state interpretation,\cite{Everett} whereupon Heisenberg entered the fray, arguing that the Copenhagen interpretation was the only viable interpretation.\cite{Heisenberg_PPc3c8} He thereby transformed Bohr's views into just another interpretation of a mathematically formulated theory. Historically, Bohr's reply\cite{Bohr1935} to the argument by Einstein, Podolsky, and Rosen\cite{EPR1935} was taken as a definitive refutation by the physics community. During the ``shut up and calculate'' period of the post-war years, Bohr's perspective was lost. His paper, which only treated the mathematical formalism in a footnote, is now widely seen as missing the point.

By transmogrifying a probability algorithm---the wave function or the ``state'' vector---into a bona fide physical state, adopting the eigenvalue--eigenstate link,%
\footnote{Thus formulated by Dirac:\cite{DiracPQM} ``The expression that an observable `has a particular value' for a particular state is permissible in quantum mechanics in the special case when a measurement of the observable is certain to lead to the particular value, so that the state is an eigenstate of the observable.''} and modeling the ``process'' of measurement as a two-stage affair (``pre-measurement'' followed by ``objectification''), von Neumann created what is commonly known as the measurement problem but is more appropriately called ``the disaster of objectification''.\cite{vF1990} This is how quantum mechanics came to be labeled as ``the great scandal of physics'',\cite{Wallace2008} as a theory that ``makes absolutely no sense'',\cite{Penrose86} and as ``the silliest'' of all the theories proposed in the 20th century.\cite{Kaku95} 

What is responsible for these mischaracterizations should not be hard to detect: think of a quantum state's dependence on time as the time-dependence of an evolving physical state, rather than as the dependence of probabilities on the time of the measurement to the possible outcomes of which they are assigned, and you have two modes of evolution whereas in reality there is not even one. 

Nevertheless, today the reasons for these mischaracterizations \emph{are} hard to detect. One of these reasons is that whereas a junior-level classical mechanics course devotes a considerable amount of time to different formulations of classical mechanics, even graduate-level courses often emphasize one particular formulation of quantum mechanics almost to the exclusion of all variants, of which there are (at least) nine.\cite{Styeretal} It would seem reasonable to expect that an interpretation of quantum mechanics be based on features that are common to all formulations of the theory, not on the mathematical idiosyncrasies of a particular formulation, such as the wave-function formulation. What is common to all formulations is that they afford tools for calculating correlations between measurement outcomes.

Another reason is the axiomatic method by which quantum mechanics is now typically taught. First students are told that the state of a quantum-physical system is (or is represented by) a normalized element of a Hilbert space. Then they are told that observables are (or are represented by) self-adjoint operators, and that the possible outcomes of a measurement are the eigenvalues of such an operator. Then comes a couple of axioms concerning the time evolution of states---unitary \emph{between} measurements and as stipulated by the projection postulate \emph{at the time of} a measurement. A further axiom stipulates that the states of composite systems are (or are represented by) vectors in the tensor product of the Hilbert spaces of the component systems. And finally, almost as an afterthought, comes an axiom about probabilities, the Born rule. This is how the \emph{The Ashgate Companion to Contemporary Philosophy of Physics}\cite{Wallace2008} comes to distinguish between a ``bare quantum formalism,'' which it describes as ``an elegant piece of mathematics'' that is  ``prior to any notion of probability, measurement etc.,'' and a ``quantum algorithm,'' which it describes as ``an ill-defined and unattractive mess,'' whose business is to extract ``empirical results'' from the former. In actual fact, there is no such thing as a bare quantum formalism. Every single axiom of the theory only makes sense as a feature of a probability calculus.\cite{Mohrhoff-QMexplained}

It is beyond doubt that significant progress was made during the roughly four decades between the passing of Niels Bohr and the advent of QBism. We now have a congeries of complex, sophisticated, and astonishingly accurate probability algorithms---the standard model%
\footnote{ ``Standard model is a grotesquely modest name for one of humankind's greatest achievements''.\cite{Wilczek2008}}%
---and we are witnessing rapid growth in the exciting fields of quantum information and quantum technology. By contrast, the contemporaneous progress in quantum theory's philosophical foundations mainly consisted in finding out what does \emph{not} work, such as the countless attempts to transmogrify statistical correlations between observations into physical processes that take place between and give rise to observations.

\section{QBism: Wigner's friend}\label{ujmsec:qbismWigner}
To make the centrality of human experience duly and truly stick, QBism emphasizes the \emph{individual} subject. To a QBist, all probabilities are of the subjective, personalist Bayesian kind. The so-called quantum state is something the individual user (of quantum mechanics) or agent (in a quantum world) assigns on the basis of her own experiences,%
\footnote{While Fuchs and Schack prefer the term ``agent,'' Mermin prefers the term ``user,'' in order to emphasize that QBists regard quantum mechanics as a ``user's manual''.\cite{MerminQBnotCop}}
and it is used by her to assign probabilities to a set of possible personal experiences, which are determined by the action she takes to elicit one these experiences. Such an action does not have to take place in a physics laboratory. It ``can be anything from running across the street at L'\'{E}toile in Paris (and gambling upon one's life) to a sophisticated quantum information experiment (and gambling on the violation of a Bell inequality)''.\cite{Fuchs_Notwithstanding} The only thing a QBist ``does not model with quantum mechanics is her own direct internal awareness of her own private {experience}''.\cite{FMS2014}

Two of the pseudo-problems that quantum-state realists have to contend with are thereby taken care of: the matter of Wigner's friend\cite{Wigner61} and the matter of Bell's shifty split.\cite{Bell90} In Wigner's scenario, Wigner's friend~$F$ performs a measurement on a system~$S$ using an apparatus~$A$. Treating $F$ as a quantum system, and treating quantum states as ontic states evolving unitarily between measurement-induced state reductions, Wigner concludes that a reduction of the combined system $S{+}A$ occurs for $F$ when she becomes aware of the outcome, while a reduction of  the combined system $S{+}A{+}F$ occurs for him when he is informed of the outcome by~$F$. This scenario led Wigner to conclude that the theory of measurement was logically consistent only ``so long as I maintain my privileged position as ultimate observer.'' QBism, on the contrary, maintains that Wigner's state assignment, which is based on his actual past and possible future experiences, is as valid as his friend's, based as that is on a different set of actual past and possible future experiences. This point, however, can be made without envisioning Wigner's friend in a coherent superposition of two distinct cognitive states:
\begin{quote}
Wigner's quantum-state assignment and unitary evolution for the compound system are only about his \emph{own} expectations for his \emph{own} experiences should he take one or another action upon the system or any part of it. One such action might be his sounding the verbal question, ``Hey friend, what did you see?,'' which will lead to one of two possible experiences for him. Another such action could be to put the whole conceptual box into some amazing quantum interference experiment, which would lead to one of two completely different experiences for him. \cite{Fuchs_Notwithstanding} 
\end{quote}
QBists distinguish between (subjective) agent-dependent realities and a common body of (objective) reality:
\begin{quote}
What is real for an agent rests entirely on what that agent experiences, and different agents have different experiences. An agent-dependent reality is constrained by the fact that different agents can communicate their experience to each other, limited only by the extent that personal experience can be expressed in ordinary language\dots. In this way a common body of reality can be constructed.\cite{FMS2014}
\end{quote}
What do we know about our common body of reality? Because we construct it from our experiences, and because our experiences are definite and irreversible, it is constructed from experiences that are definite and irreversible. I may be ignorant of your experiences and you may be ignorant of mine, but we cannot doubt the definiteness and irreversibility of our respective experiences. It is therefore inadmissible to assign to any (sane and healthy) subject a coherent superposition of distinct cognitive states. Wigner is not only perfectly justified but \emph{required} to assign to the system that includes his friend an incoherent mixture reflecting his ignorance of the outcome that his friend has obtained. To treat his own experiences as definite but not those of his friend---that would be the solipsism which Wigner feared and sought to avoid by proposing ``that the equations of motion of quantum mechanics cease to be linear, in fact that they are grossly non-linear if conscious beings enter the picture.''

QBists are united in rejecting ``the silly charges of solipsism''.\cite{MerminQBnotCop} In order to avoid these charges, however, they need to do more than acknowledge the fact that ``[m]y experience of you leads me to hypothesize that you are a being very much like myself, with your own private experience.'' They need to stop fantasizing about coherent superpositions involving distinct experiences.

\section{QBism: Bell's shifty split}\label{ujmsec:qbismShifty}
{\leftskip\parindent\emph{There is a straight ladder from the atom to the grain of sand, and the only real mystery in physics is the missing rung. Below it, particle physics; above it, classical physics; but in between, metaphysics.}\par\hfill--- \emph{Tom Stoppard, \emph{Hapgood}}}\par\smallskip\noindent
Bell's ``shifty split,'' a.k.a. the Heisenberg cut, is the mysterious boundary separating the system under investigation from the means of investigation. While for Heisenberg its location was more or less arbitrary, for Bohr it was determined by the measurement setup.%
\footnote{See Camilleri and Schlosshauer\cite{CamilleriSchlosshauer2015} for a discussion of Bohr's and Heisenberg's divergent views on this matter.}
In QBism, the experience of the individual user takes the place of the measurement setup. Accordingly there are as many splits as there are users, and there is nothing shifty about them. Mermin explains:
\begin{quote}
Each split is between an object (the world) and a subject (an agent's irreducible awareness of her or his own experience). Setting aside dreams or hallucinations, I, as agent, have no trouble making such a distinction, and I assume that you don't either. Vagueness and ambiguity only arise if one fails to acknowledge that the splits reside not in the objective world, but at the boundaries between that world and the experiences of the various agents who use quantum mechanics.\cite{Mermin_shifty}
\end{quote}
Let us disregard the ambiguity of ``awareness of one's own experience,'' which could mean either awareness of something one is experiencing or awareness of one's experiencing something. The question is: what is meant by ``objective world''? First guess: what is meant is ``the common external world we have all negotiated with each other'' or, equivalently, ``a model for what is common to all of our privately constructed external worlds''.\cite{MerminQBnotCop} In this case the split occurs between this world or model and the private experiences in which it originates. 

Second guess: what is meant is something that induces experiences in conscious subjects. This interpretation is suggested by Mermin's statement that, according to QBism, ``my understanding of the world rests entirely on the experiences that the world has induced in me throughout the course of my life'',\cite{MerminQBnotCop} or by the equivalent statement that  ``[t]he world acts on me, inducing the private experiences out of which I build my understanding of my own world''.\cite{MerminBetterSense} Judging by \emph{these} statements, the split occurs between my private experiences and a world that induces them, rather than between my private experiences and the world as I understand it on the basis of my private experiences. (On the relation between these two worlds see Note~\ref{note:sacop}.)

What we are faced with here is an attempt to throw the baby out with the bathwater. The bathwater is the shifty split; the baby is the measuring apparatus. And not only the measuring apparatus. If QBism, as Fuchs and Schack affirm, treats ``all physical systems in the same way, including atoms, beam splitters, Stern-Gerlach magnets, preparation devices, measurement apparatuses, all the way to living beings and other agents'',\cite{FS2015} then Bohr's crucial insight that the properties of quantum systems are \emph{contextual}---that they are defined by experimental arrangements---is lost.  

For Bohr, the measurement apparatus was needed not only to indicate the possession of a property (by a system) or a value (by an observable) but also, and in the first place, to make a set of properties or values available for attribution to a system or an observable. The sensitive regions of an array of detectors \emph{define} the regions of space in which the system can be found. In the absence of an array of detectors, the regions of space in which the system can be found do not exist. The orientation of a Stern-Gerlach apparatus \emph{defines} the axis with respect to which a spin component is measured.  In the absence of a Stern-Gerlach apparatus, the axis with respect to which a spin component can be up or down does not exist. What physical quantity is defined by running across the street at L'\'{E}toile in Paris?

From a different QBist point of view, espoused by Fuchs and Schack, the measurement apparatus should be understood as an extension of the agent, and quantum mechanics itself should be regarded as a theory of stimulation and response: 
\begin{quote}
A quantum measurement device is like a prosthetic hand, and the outcome of a measurement is an unpredictable, undetermined ``experience'' shared between the agent and the external system.\cite{Fuchs_Notwithstanding}
\end{quote}\begin{quote}
The agent, through the process of quantum measurement stimulates the world external to himself. The world, in return, stimulates a response in the agent that is quantified by a change in his beliefs---i.e., by a change from a prior to a posterior quantum state. Somewhere in the structure of those belief changes lies quantum theory's most direct statement about what we believe of the world as it is without agents.\cite{FS2004}
\end{quote}
This invites two comments. The first is that the question where the apparatus ends and the rest of the world begins is once more open to dispute. It appears that one shifty split has been traded for another. Fuchs responds by pointing out that the physical extent of the agent is up to the agent:
\begin{quote}
The question is not where does the quantum world play out and the classical world kick in? But where does the agent expect his own autonomy to play out and the external world, with its autonomy and its capacity to surprise, kick in? The physical extent of the agent is a judgment he makes of himself. \cite{Fuchs2Wootters}
\end{quote}
By placing the the dividing line---wherever the agent chooses to place it---between the agent-cum-instrument and the rest of the physical world, Fuchs does precisely what Mermin objects to when he writes that ``[v]agueness and ambiguity only arise if one fails to acknowledge that the splits reside not in the objective world, but at the boundaries between that world and the experiences of the various agents who use quantum mechanics''.\cite{Mermin_shifty}

The second comment concerns ``the world as it is without agents.'' The phrase \emph{could} refer to the unspeakable domain beyond the reach of our concepts, which only becomes speakable through the manner in which it is stimulated (i.e., by saying in ordinary language what the agent has done) and through the manner in which it responds (i.e., by saying in ordinary language what the agent has learned). Mermin,\cite{Mermin2ujm} however, rejects this interpretation: ``QBists (at least this one) attach no meaning to `the world as it is without agents.' It only means `the common external world we have all negotiated with each other'.''

Regardless, there is another boundary at which the Heisenberg cut can be placed. As our common external world has a ``near'' boundary (between it and the private experiences in which it originates), so it has a ``far'' boundary (between it and the unspeakable domain beyond the reach of our concepts). My contention is that the cut ought to be placed there. I therefore agree with Mermin that the cut does not reside in the objective world---the world of sense impressions organized into objects, the world we have all negotiated with each other. But instead of placing it at its near boundary, I maintain that it should be placed at its far boundary, and I take it that, to all intents and purposes, Bohr did the same. What is definite in this case is not just the measurement apparatus but the entire objective world. Nothing indefinite is implied by the unpredictability in general of measurement outcomes.

\section{QBism and Bohr}\label{ujmsec:QBB}
As it is an irony that Bohr drew the battle lines in a way which put Kant and himself
on opposing sides, so it is an irony that QBists draw their battle lines in a way which puts Bohr and themselves on opposing sides, notwithstanding that ``QBism agrees with Bohr that the primitive concept of \emph{experience} is fundamental to an understanding of science''.\cite{FMS2014} 
Thus Fuchs \emph{et al.}:
\begin{quote}
The Founders of quantum mechanics were already aware that there was a problem. Bohr and Heisenberg dealt with it by emphasizing the inseparability of the phenomena from the instruments we devised to investigate them. Instruments are the Copenhagen surrogate for experience\dots. [They are] objective and independent of the agent using them.\cite{FMS2014}
\end{quote}
And thus Mermin: 
\begin{quote}
Those who reject QBism \dots\ reify the common external world we have all negotiated with each other, purging from the story any reference to the origins of our common world in the private experiences we try to share with each other through language. \dots by ``experience'' I believe [Bohr] meant the objective readings of large classical instruments\dots. Because outcomes of Copenhagen measurements are ``classical,'' they are \emph{ipso facto} real and objective.\cite{MerminQBnotCop}
\end{quote}
While QBists are generally aware of the important distinction between what Mermin calls ``reification'' and what Schr\"odinger\cite{SchrWhatIsReal} called ``objectivation,'' they share the now prevailing misappreciation of Bohr's thinking. Bohr was concerned with objectivation, the representation of a shared mental construct as an objective world, not with reification, which ignores or denies the origins of the objective world in our thoughts and perceptions. Objectivation means purging from the story any reference to these origins without ignoring or denying them, so that science may deal with the objective world as common-sense realism does---\emph{as if} it existed independently of our thoughts and perceptions. Reification is the assertion that the world we perceive does in fact exist independently of our perceptions, or that the world we mentally construct does in fact exist independently of our constructing minds, or that the world we describe does in fact exist---just as we describe it---independently of our descriptions.

To Bohr, measurement outcomes are ``classical'' (i.e., definite and irreversible) and instruments are objective not because (or in the sense that) they are reified but because (or in the sense that) they are situated in an intersubjectively constituted world---like everything else that is directly accessible to human sensory experience. Instead of being a ``surrogate of experience,'' instruments---like everything else in our common external world---are experiences that lend themselves to objectivation. They make it possible not only to apply classical concepts to quantum systems but also to extend their reach into the non-classical domain via principles of correspondence.%
\footnote{``[Q]uantum mechanics and quantum field theory only refer to individual systems due to the ways in which the quantum models of matter and subatomic interactions are linked by semi-classical models to the classical models of subatomic structure and scattering processes. All these links are based on tacit use of a generalized correspondence principle in Bohr's sense (plus other unifying principles of physics).''  This generalized correspondence principle serves as ``a semantic principle of continuity which guarantees that the predicates for physical properties such as `position', `momentum', `mass', `energy', etc., can also be defined in the domain of quantum mechanics, and that one may interpret them operationally in accordance with classical measurement methods. It provides a great many inter-theoretical relations, by means of which the formal concepts and models of quantum mechanics can be filled with physical meaning''.\cite{Falkenburg2007-XII-191}}

The statement that those who reject QBism reify the common external world we have all negotiated with each other, also rings false. One can certainly reject some of the (sometimes mutually inconsistent) claims made by QBists without reifying the objectivized world. What is true is the converse: those who reify the objectivized world will have to reject QBism.

Admittedly, Bohr obscured his original thinking by compounding the incontestable irreversibility of human sensory experience with amplification processes or  apparatus features like being sufficiently large or heavy. To invoke such processes or features was for him the price that one had to pay for achieving complete objectivation (i.e., for banishing every reference to experience). He himself, however, was clearly not inclined to pay this price. It bears repetition: for him the description of atomic phenomena had ``a perfectly objective character, \emph{in the sense that} no explicit reference is made to any individual observer and that therefore \dots\ no ambiguity is involved in the communication of information'' [BCW\,7:390, emphasis added]---and thus \emph{not in the sense that} no reference was made to the community of communicating observers or to the incontestable irreversibility of their experiences. 

To Bohr, objectivity meant ``a description by means of a language common to all \dots\ in which people may communicate with each other in the relevant field'' [BCW\,10:XXXVII]. What QBists mean by objectivity is less clear, though arguably they mean the same, to wit: ``language is the only means by which different users of quantum mechanics can attempt to compare their own private experiences,'' and it is only by communicating ``that we can arrive at a shared understanding of what is common to all our own experiences of our own external worlds''.\cite{MerminQBnotCop}

There is just one detail in Mermin's argument to which Bohr would probably have objected, namely the idea that each of us first constructs a private external world, and that language comes in only after this is done, as a means of figuring out what is common to all our privately constructed external worlds. One cannot construct a private external world before being in possession of a language providing the concepts that are needed for its construction. At any rate, Mermin's claim that  ``[o]rdinary language comes into the QBist story in a more crucial way than it comes into the story told by Bohr,'' appears to me wholly unjustified.

The great merit of QBism is that it puts the spotlight back on the role that human experience plays in creating physical theories. One this is recognized, the mystery as to why measurements are irreversible and outcomes definite vanishes into this air: it is because our experiences are irreversible and definite. Bohr could have said the same, and arguably did, but in so many words that the core of his message has been lost or distorted beyond recognition. The fundamental difference between Bohr and QBism is that the former was writing before interpreting quantum mechanics became a growth industry, while the latter emerged in reaction to an ever-growing number of futile attempts at averting the disaster of objectification in the same realist framework in which it arose.

There are several ways in which QBism goes beyond Bohr, but this in no wise affects my claim that Bohr was a QBist---what with his contention that ``in our description of nature the purpose is not to disclose the real essence of the phenomena but only to track down, so far as it is possible, relations between the manifold aspects of our experience'' [BCW\,6:296]. And no, by ``our experience'' he did \emph{not} mean the objective readings of large classical instruments. What qualifies him as a QBist, if not as the unwitting \emph{founder} of QBism, is his insistence on \emph{individual} experience, which only through communication becomes \emph{objective} knowledge.

One of the ways in which contemporary QBism goes beyond Bohr consists in replacing the standard projector-valued probability measures by positive-operator-valued measures (POVMs). This makes it possible to formulate the Born rule entirely in terms of probabilities (as outlined in the Appendix) and to view quantum mechanics as nothing but a generalization of the (personalist) Bayesian theory of probability. In their justifiable enthusiasm, however, QBists also overshoot their mark, as when they permit incoherent superpositions to be assigned to possibilities involving distinct cognitive states, or when they claim that quantum mechanics is ``\emph{explicitly} local in the QBist interpretation''.\cite{FMS2014} According to Fuchs,\cite{Fuchs2010} 
\begin{quote}
[QBism] gives each quantum state a home. Indeed, a home localized in space and time---namely, the physical site of the agent who assigns it! By this method, one expels once and for all the fear that quantum mechanics leads to ``spooky action at a distance.''
\end{quote}
A quantum state has its home in an agent's mind, not at any physical site---which would have to be a site in our common external world, since there are no agents in ``the world as it is without agents.'' Why inserting minds into our common external world is a bad idea has been explained by Schr\"odinger, who according to Fuchs \emph{et al.} took ``a QBist view'' of science.\cite{FMS2014} By the very fact that we treat some of our experiences as aspects of a shared external world, Schr\"odinger argued, we exclude ourselves from this world: ``We step with our own person back into the part of an onlooker who does not belong to the world, which by this very procedure becomes an objective world''.\cite{SchrLifeMindMatter} If we then reify the mind's creation, we are left with no choice but to insert the mind into its creation: ``I so to speak put my own sentient self (which had constructed this world as a mental product) back into it---with the {pandemonium} of disastrous logical consequences'' that flow from this error, such as ``our fruitless quest for the place where mind acts on matter or vice-versa.'' Locating experiences in our common external world therefore is not an option. Instead of asserting that QBism is explicitly local, QBists ought to assert that QBism is neither local nor nonlocal in any realist sense of these terms.

It is strange indeed to see a QBist look upon spooky action at a distance as something to be feared. To banish it by claiming that quantum mechanics is local is to concede way too much to those who see themselves as modeling a reality not of their own making. Something fearsome is implied only if one forgets that physics deals with our common external world. One then has to worry how, in the absence of a common cause, measurement outcomes in spacelike relation can be so spookily correlated. Keeping in mind that measurement outcomes are responses from a reality beyond the reach of human sensory experience, one realizes (as Bohr did) that the answer to this question is beyond the reach of our concepts. After all, diachronic correlations (between successive experiences of the same agent) are no less inexplicable than synchronic correlations (between simultaneous experiences of different agents). 

All in all, QBism, through its emphasis on the individual experiencing subject, brings home the intersubjective constitution of our common external world more forcefully than Bohr ever did. (The time wasn't ripe for this then. Perhaps it is now.) Bohr's insights, on the other hand, are eminently useful in clarifying the QBist position, attenuating its excesses, and enhancing its internal consistency.

In concluding, I want to extend my gratitude to the QBists for making me finally come round to seeing that there is \emph{no} difference between observations qua experiences and observations qua measurement outcomes (which explains why measurements are irreversible and outcomes definite). 

\section*{Appendix: The Born rule according to QBism}\label{ujmapx}
{\leftskip2\parindent\emph{If quantum theory is so closely allied with probability theory, why is it not written in a language that \emph{starts} with probability, rather than a language that ends with it? Why does quantum theory invoke the mathematical apparatus of Hilbert spaces and linear operators, rather than probabilities outright? --- Christopher A. Fuchs\cite{Fuchs2010}}\par}\smallskip\noindent 
For QBists, quantum mechanics is a generalization of the Bayesian theory of probability. It is a calculus of consistency---a set of criteria for testing coherence between beliefs. As there are no {external} criteria for declaring a probability judgment right or wrong, so there are no {external} criteria for declaring a quantum state assignment right or wrong. The only criterion for the adequacy of a probability judgment or a state assignment is internal coherence between beliefs. The Born rule thus is not simply a rule for updating probabilities, for getting new ones from old. It is a rule for {relating} probability assignments and {constraining} them. As such, it can be expressed entirely in terms of probabilities. 

The proof of the last claim requires the use of POVMs, which generalize the standard projector valued measures used by von Neumann\cite{vN1932} and Dirac.\cite{DiracPQM} It goes like this: While a density operator $\rho$ determines a potentially infinite number of probabilities, these cannot all be independent. On a $d$-dimensional Hilbert space, $\rho$~is completely determined by the $d^{\,2}$ probabilities it assigns to the outcomes (represented by linearly independent positive operators~$E_i$) of an \emph{informationally complete} measurement. Any density operator $\rho$ therefore corresponds to a vector whose $d^{\,2}$ components are the Born probabilities
\begin{equation}
p_i=\hbox{Tr}(\rho E_i)\,,
\end{equation}
and any POVM $\{F_j\}$ corresponds to a matrix whose elements are the conditional probabilities
\begin{equation}
R_{ji}=\hbox{Tr}(\Pi_i F_j)\,,
\end{equation}
where the $\Pi_i$ are 1-dimensional projectors proportional to~$E_i$. This makes it possible to write the Born rule in the generic form
\begin{equation}
q(F_j)= f \bigl(\{R_{ji}\},\{p_i\}\bigr)\,,
\label{eq:BornGeneric}
\end{equation}
where $f$ depends on the details of the informationally complete measurement $\{E_i\}$.

The function $f$ takes a particularly simple form if the positive operators $E_i$ constitute a \emph{symmetric} informationally complete (SIC) measurement. In this case one of the ways in which the Born rule can be written is\cite{Fuchs2010, Fuchs2004}
\begin{equation}
q(F_j)=\sum_{i=1}^{d^2}\left[(d+1)\,p_i-\frac1d\right]R_{ji}\,.
\label{eq:BornPOVM}
\end{equation}
If the positive operators $F_j$ are mutually orthogonal projectors representing the outcomes of a complete von Neumann measurement, the Born rule takes the even simpler form
\begin{equation}
q(F_j)=(d+1)\sum_{i=1}^{d^2}\,R_{ji}p_i-1\,.
\label{eq:BornPVM}
\end{equation}
While the probabilities (\ref{eq:BornPOVM}) and (\ref{eq:BornPVM}) are expressed in terms of (i)~the probabilities $p_i$ that an agent assigns to the possible outcomes of the SIC measurement and (ii)~the conditional probabilities $R_{ji}$ that the agent assigns to the possible outcomes of a subsequent measurement if the SIC measurement is actually made, they pertain to a situation in which the SIC measurement is \emph{not} made. If it \emph{is} made, the law of total probability applies, and we have
\begin{equation}
q(F_j)=\sum_{i=1}^{d^2}R_{ji}p_i \,.
\label{eq:BornAgain}
\end{equation}
Comparing Eqs. (\ref{eq:BornPOVM}) and (\ref{eq:BornPVM}) with Eq. (\ref{eq:BornAgain}), one can see that ``[t]he Born Rule is nothing but a kind of Quantum Law of Total Probability!  No complex amplitudes, no operators---only probabilities in, and probabilities out''.\cite{Fuchs2010}

QBists hope to eventually be in a position to derive the standard Hilbert space formalism from the Born rule. And they hope so to distill the essence of quantum mechanics and the essential characteristic of the quantum world.%
\footnote{While the Born rule is normative---it guides an agent's behavior in a world that is fundamentally quantum---it is also an empirical rule. It is a statement about the quantum world, indirectly expressed as a calculus of consistency for bets placed on the outcomes of measurements.}
This a fascinating, highly ambitious, and  seriously challenging project. Do SIC measurements even exist? Unfortunately, proofs of their existence are elusive. As of May 2017, such proofs have been found for all dimensions up to $d{=}151$, and for a few others up to 323.\cite{Fuchs_Notwithstanding} The mood of the QBist community nevertheless is that a SIC measurement should exist for every finite dimension. That said, it must be stressed that the general form of the Born rule, Eq.~(\ref{eq:BornGeneric}), does not depend on the existence of SIC measurements; it only presupposes informationally complete POVMs, and these are known to exist for all finite dimensions.

\end{document}